\documentclass[useAMS,usenatbib]{mn2e}
\usepackage{color}
\usepackage{graphicx} 
\usepackage{subfigure} 
\usepackage{amsmath} 
\usepackage{rotating}
\usepackage{multicol}
\usepackage{multirow}
\usepackage[colorlinks=true,citecolor=blue]{hyperref} 
\usepackage[utf8]{inputenc} 
\usepackage[normalem]{ulem} 

\def\mjy{\,\ensuremath{\mathrm{mJy}}}
\def\ghz{\,\ensuremath{\mathrm{GHz}}}
\def\mhz{\,\ensuremath{\mathrm{MHz}}}

\def\degsq{\,{\ensuremath{\mathrm{deg}^{-2}}}}
\def\sqdeg{\,{\ensuremath{\mathrm{deg}^{2}}}}

\defcitealias{thyagarajan_variable_2011}{T11}
\defcitealias{Hodge_millijansky_2013}{H13}

\title[Variability in the PDS]{Radio variability in the Phoenix Deep Survey at 1.4GHz}
\author[Hancock et al.]{P. J. Hancock$^{1,2,3}$\thanks{E-mail:
Paul.Hancock@Curtin.edu.au}, J. A. Drury$^{2}$, M. E. Bell$^{2,3,4}$, T. Murphy$^{2,3}$, B. M. Gaensler$^{2,3,5}$\\
$^{1}$ International Centre for Radio Astronomy Research, Curtin University, Bentley, WA 6102, Australia\\
$^{2}$ Sydney Institute for Astronomy (SIfA), School of Physics, The University of Sydney, Australia 2006\\
$^{3}$ ARC Centre of Excellence for All-sky Astrophysics (CAASTRO)\\
$^{4}$ CSIRO Astronomy and Space Science, PO Box 76, Epping, NSW 1710, Australia\\
$^{5}$ Dunlap Institute for Astronomy \& Astrophysics, University of Toronto, Ontario, Canada}
\begin{document}

\date{}

\pagerange{\pageref{firstpage}--\pageref{lastpage}} \pubyear{2016}

\maketitle

\label{firstpage}

\begin{abstract}

We use archival data from the Phoenix Deep Survey to investigate the variable radio source population above 1\mjy{}/beam at 1.4\ghz{}.
Given the similarity of this survey to other such surveys we take the opportunity to investigate the conflicting results which have appeared in the literature. 
Two previous surveys for variability conducted with the Very Large Array (VLA) achieved a sensitivity of 1\mjy/beam.
However, one survey found an areal density of radio variables on timescales of decades that is a factor of $\sim 4$ times greater than a second survey which was conducted on timescales of less than a few years. 
In the Phoenix deep field we measure the density of variable radio sources to be $\rho=0.98\mathrm{deg}^{-2}$ on timescales of 6 months to 8 years. 
We make use of WISE infrared cross-ids, and identify all variable sources as an AGN of some description. 
We suggest that the discrepancy between previous VLA results is due to the different time scales probed by each of the surveys, and that radio variability at 1.4\ghz{} is greatest on timescales of $2-5$ years.

\end{abstract}


\section{Introduction} 

There are many mechanisms which can cause variability in radio sources both
intrinsic and extrinsic. Intrinsic variability can be due to variable
accretion rates onto black holes, flare like activity in stars, tidal
disruption events, or explosive events such as novae and supernovae. Extrinsic
variability is typically induced either by scintillation due to turbulence
in the interplanetary or interstellar medium (or the ionosphere at very low
frequencies), or by extreme scattering events \citep{Fiedler_summary_1994}.
Each of these mechanisms is understood, however the relative incidence and
magnitude is less well understood. 

The incidence of radio variability has been investigated with the aid of blind
surveys using a combination of new and archival data. A common metric that is
used to compare surveys with different attributes is the two epoch equivalent
source density \citep{bower_submillijansky_2007}. Over time a trend has
emerged: the more sensitive the survey detection limit, the greater the number
of variable sources that are detected \citep{Mooley_caltech-nrao_2016}. This
trend is expected as  the source count distribution of radio sources - there
are more faint sources than bright sources. However there are many additional
factors that contribute to differences in the density of variable radio
sources including: observing frequency, observing cadence, integration time,
and galactic latitude.

As the amount of intervening gas from the Milky Way's interstellar medium
(ISM) increases with lines of sight with lower Galactic latitudes, there is
expected to be an increase in interstellar scintillation towards the Galactic
plane. \citet{Gaensler_long-term_2000} found that there is a tendency for
radio sources with $|b|<20\deg$ to be more likely to be variable, but that
there was no correlation between the incidence, magnitude, or time-scale of
variability above $|b|=20\deg$. This result is supported by
\citet{ofek_structure_2011} who see a doubling in the fraction of variable
sources below  $|b|=20\deg$. \citet{Ghosh_lowfrequency_1992} find an increase
in the magnitude of variability at $10\deg<b<30\deg$ which is twice that at
both higher {\em and} lower latitudes. At latitudes above $|b|=30\deg$ there
is no indication that there is a relation between Galactic latitude and the
incidence or magnitude of variability.

In the last decade considerable effort has been spent trying to map out the
parameter space of radio variability. Much of this work has been focused
around 1\ghz{} \citep[see][and references there in]{Mooley_caltech-nrao_2016},
however studies at higher frequencies have also been done \citep[eg,
][]{bell_search_2015,bower_submillijansky_2007}. The studies at 1.4\ghz{} have
spanned at least four orders of magnitude in sensitivity, giving a fairly
sparse coverage of the parameter space. To date there have been only two blind
surveys that have probed the same region of parameter space close enough to
invite direct comparison, and they find a density of variable radio sources
that differ by a factor of $\sim 4$. The first survey was conducted by
\citet[][ hereafter
\citetalias{thyagarajan_variable_2011}]{thyagarajan_variable_2011}, who
compared the NVSS and FIRST survey images to search for radio variability,
found variable sources with a density of $\rho=0.2\mathrm{deg}^{-2}$ above
1\mjy/beam. The second survey was conducted by \citet[][ hereafter
\citetalias{Hodge_millijansky_2013}]{Hodge_millijansky_2013}, who compared
multiple epochs of the FIRST survey and a new set of VLA observations, and
found variable sources with a density of  $\rho=0.74\mathrm{deg}^{-2}$ also
above 1\mjy/beam. The two surveys used the same telescope and frequency, had
the same sensitivity, and both used data from the FIRST survey, and yet
arrived at significantly different results. This discrepancy has not yet been
explored in the literature. We therefore focus on the difference between the
two surveys in order to explain the differing results.

In this paper we use archival data from the Phoenix Deep Survey to conduct a
blind search for variable radio sources at 1.4\ghz{}. The survey achieves a
$5\sigma$ sensitivity of 1\mjy/beam and can thus be used to understand the
conflicting  results of \citetalias{thyagarajan_variable_2011} and
\citetalias{Hodge_millijansky_2013}.

In section~\ref{sec:comparison} we compare and contrast the
\citetalias{thyagarajan_variable_2011} and \citetalias{Hodge_millijansky_2013}
surveys and motivate the work of this paper. In section~\ref{sec:data} we
detail the data acquisition and reduction. In
sections~\ref{sec:image_analysis}-\ref{sec:lc_analysis} we present the image
and light curve analysis. We discuss the variables sources in
section~\ref{sec:variable_source_analysis}, and our results in
section~\ref{sec:discussion}. We summarize and draw conclusions in
section~\ref{sec:summary}.

\section{Survey comparison}\label{sec:comparison}

\citetalias{thyagarajan_variable_2011} used data from the FIRST survey of the
northern Galactic cap, covering nearly $8500\,\mathrm{deg}^2$. The survey area
includes Galactic latitudes from $+17\deg$ to $+90\deg$, with 90\% of the
variable sources found above a latitude of $+30\deg$.
\citetalias{Hodge_millijansky_2013} used data from the FIRST survey, and
additional follow-up observations, to survey the SDSS stripe 82 region
covering $60~\mathrm{deg}^2$. The area surveyed by
\citetalias{Hodge_millijansky_2013} is restricted to a Galactic latitude of
$40\deg<b<45\deg$. \citetalias{thyagarajan_variable_2011} and
\citetalias{Hodge_millijansky_2013} survey different areas of sky but $>90\%$
of sources in these two surveys lie above $b=30\deg$. Since this is outside
the $10\deg<b<20\deg$ enhancement region identified by
\citep{Ghosh_lowfrequency_1992} and others, Galactic latitude effects cannot
be responsible for the different source densities observed.

Another difference between the \citetalias{thyagarajan_variable_2011} and
\citetalias{Hodge_millijansky_2013} surveys is the time scales of variability
that are probed. \citetalias{thyagarajan_variable_2011} worked with images that
were separated by as little as 3~minutes up to a few years, with the majority
of differences necessarily being at short time scales.
\citetalias{Hodge_millijansky_2013} worked with three images each separated by
7~years. Thus \citetalias{thyagarajan_variable_2011} is more sensitive to
short term variability, whilst \citetalias{Hodge_millijansky_2013} is
sensitive only to long term variability on 7 year timescales.
\citet{ofek_structure_2011} compared fluxes measured in the NVSS and FIRST
surveys and showed that the amount of variability doesn't change with
observing cadences between $2-5$ years. In this work we use data with
a cadence of between 150 and 2000 days, falling right between the peak
sensitivity of the \citetalias{thyagarajan_variable_2011} and
\citetalias{Hodge_millijansky_2013} surveys. In a study at 843\mhz,
\citet{bannister_22-yr_2011} found some evidence that there is a peak in radio
variability on timescales of between 2000 and 3000 days, which would suggest
that the increased detection rate of \citetalias{Hodge_millijansky_2013} is a
feature of the mechanism that is causing the radio variability.

The Phoenix Deep Survey microjansky catalog
\citep[PDS,][]{hopkins_phoenix_2003} was constructed from six epochs of data
taken with the Australia Telescope Compact Array (ATCA) at 1.4\ghz{} over a
period of 8 years. The mosaicked images from each epoch of observations
achieve a sensitivity of $\sim1$\mjy/beam. We use the similarity between the
frequency and sensitivity of these observations and the
\citetalias{thyagarajan_variable_2011} and \citetalias{Hodge_millijansky_2013}
observations, to investigate the conflicting results. The PDS data have a
cadence that is between that of \citetalias{thyagarajan_variable_2011} and
\citetalias{Hodge_millijansky_2013} and also covers the peak suggested by
\citet{bannister_22-yr_2011}. We can therefore determine whether observing
cadence plays an important role in the rate of variable sources that are
observed in a given survey.

\section{Data Acquisition and Reduction}\label{sec:data}

We use archival observations of the PDS observed with the Australia Telescope
Compact Array between 1993 and 2001 as part of the Phoenix Deep survey
\citep[PDS,][]{hopkins_phoenix_2003}. The combined observations cover a $5.97$\degsq{}
region of the sky at 1.4\,GHz. The region is bound by 01:05:35$<$RA$<$01:22:22
and -47:01:59$<$Dec$<$-44:25:08.

Calibrated data were obtained from the PDS group. The calibration and imaging
of these data are described in \citet{Hopkins_phoenix_1998}, and
\citet{Hopkins_microjansky_1999}. As noted by \citet{hopkins_phoenix_2003} the
Phoenix field contains a number of bright sources that are difficult to clean
completely and thus some mosaics contain artefacts around such sources. Since
side-lobes and image artefacts can masquerade as variable or transient events,
a second round of cleaning  was performed around bright sources. This second
round of cleaning significantly reduced the magnitude of the artefacts,
however these artefacts still dominated the regions around bright sources. The
data were grouped into six epochs, each with a duration of 1-22 days.
Table\,\ref{tab:epochs} shows the observing dates, total imaged area, and
median $5\sigma$ sensitivity, for each of the six epochs.

\begin{table}
\centering
\begin{tabular}{rlcc}
\hline
Observing  & Name & Area               & Sensitivity \\
Date(s) &       & ($\mathrm{deg}^2$) &  $5\sigma$(mJy) \\
\hline
28 Jan - 30 Jan 1994 & 1994E & 3.79 & 2.29 \\
 3 Jul - 6  Jul 1994 & 1994L & 4.50 & 1.82 \\
27 Nov - 18 Dec 1997 & 1997  & 2.03 & 0.84 \\
15 Sep 1999          & 1999  & 1.86 & 0.71 \\
 9 Sep- 13 Sep 2000  & 2000  & 2.62 & 0.69 \\ 
 1 Aug 2001          & 2001  & 1.59 & 0.76 \\
\hline
\end{tabular}
\caption{The observing dates, area, and median $5\sigma$ sensitivity of each epoch.}
\label{tab:epochs}
\end{table}

\section{Image Analysis}\label{sec:image_analysis}

We used the prototype pipeline developed for the Variables And Slow Transients
\citep[VAST,][]{murphy_vast:_2013} survey \citep{banyer_vast:_2012} to automate
the source finding, crossmatching, and variability analysis. As part of the
VAST pipeline, a mosaic of each epoch was processed using the Aegean source
finding algorithm \citep{Hancock_compact_2012}.

\subsection{Background and Noise characterization} 

\citet{Mooley_sensitive_2013} demonstrated that the Aegean source finding
algorithm does not perform well in regions of an image where the background or
noise is changing rapidly. The PDS field does not contain any significant
diffuse background emission, however the noise in the mosaics increases around
the edge of the image, and around bright sources. We use the Background And
Noise Estimation program (BANE\footnote{Available from
\href{http://github.com/PaulHancock/Aegean}{github.com/PaulHancock/Aegean}})
to improve the completeness and reliability of the source finding in the
presence of rapidly changing noise characteristics. BANE performs sigma
clipping on the pixel distribution in order to provide a much more accurate
measure of the background and noise properties of an image than the method
used internally by Aegean. We did not follow the method of \citet{Hopkins1998}
who exclude regions of sky around bright sources.

\subsection{Overlap Regions}

\begin{figure}
\centering
\includegraphics[width=\linewidth]{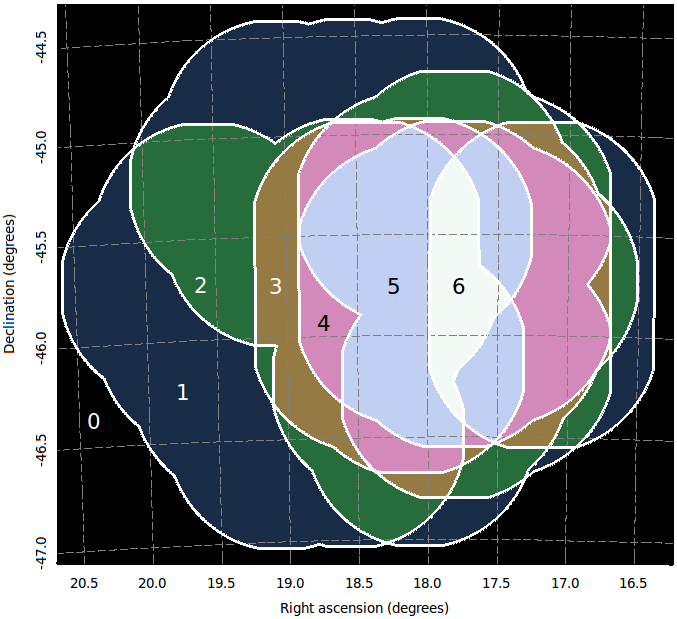}
\caption{A representation of the overlap between each of the epochs considered in this work. Each shaded region represents between 1 to 6 images covering the given area of sky, as indicated by the annotations. The area of each region is listed in column 2 of Table~\ref{tab:overlaps}.}
\label{fig:coverage}
\end{figure}

The six epochs of observations do not all cover the same area of sky. There is
a small area of sky of $0.28\,\mathrm{deg}^2$ that is imaged in all six epochs.
To increase the area available to detect variable and transient sources we
include all regions of sky that were observed in at least two epochs.
Figure\,\ref{fig:coverage} shows the area of sky that is covered by between
1-6 epochs of observations, and Table\,\ref{tab:overlaps} details the area,
median sensitivity, and number of sources detected in each overlap region.
When calculating the number of sources that are detected in each of the
overlap regions ($N_S$), a correction is made to account for the reliability
of each of the input images. For the remainder of this paper the number of
sources in an epoch $(N_S)$ indicates the number of sources detected that are
expected to be real.

\begin{table}
\centering
\begin{tabular}{cccc}
\hline
Epochs  &  Area $=N_E$ & $5\sigma$   & Sources \\
($N_E$) & (\sqdeg)     & (\mjy/beam)      & ($N_S$)  \\
\hline
 1      &   2.09       &   3.35      &   336 \\ 
 2      &   1.19       &   2.48      &   410 \\
 3      &   0.50       &   1.58      &   453 \\
 4      &   0.83       &   0.97      &   498 \\
 5      &   1.08       &   0.61      &   594 \\
 6      &   0.28       &   0.72      &   328 \\
\hline
\end{tabular}
\caption{The area, median sensitivity, and source count, for the regions depicted in Figure\,\ref{fig:coverage}. The source count ($N_S$) includes a correction for reliability.}
\label{tab:overlaps}
\end{table}

\section{Light Curve Analysis}\label{sec:lc_analysis}  

The VAST pipeline produces three metrics for
measuring the variability of a source, two metrics that measure the magnitude
of variability and one metric that measures the significance of variability.
The magnitude of variability is measured by the modulation index $m$ and the
de--biased modulation index $m_d$ as described in \citet{Bell_survey_2014} (and
references therein):

\begin{subequations}
\begin{align}
m &= \frac{\bar\sigma_s}{\bar S} \label{eq:m}\\
m_d &= \frac{1}{\bar S}\sqrt{\frac{ \sum^n_{i=1}\left(S_i-\bar S\right)^2 - \sum^n_{i=1}\sigma_i^2 }{n}},\label{eq:md}
\end{align}
\end{subequations}

\noindent where $\bar S,\bar \sigma_s$ are the mean and standard deviation of
the fluxes in a light curve,  $S_i,\sigma_i$ are individual measurements
within a light curve, and $n$ is the number of measurements. $m_d$ is taken to
be negative when the discriminant of Eq\,\ref{eq:md} is negative.

To calculate the significance of variability we first measured the
$\chi^2_{lc}$ for the light curve and computed the probability that the given
value would be seen in a non-variable source. Following
\citet{Bell_survey_2014} we calculated $\chi^2_{lc}$ as

\begin{equation}
\chi^2_{lc} = \sum^n_{i=1}\frac{(S_i-\bar{S})^2}{\sigma_i^2},
\label{eq:chi}
\end{equation}

\noindent where $S_i$ is the $i$th flux density measurement with variance
$\sigma_i^2$, and $\bar{S}$ is the weighted mean flux density. The
significance of variability indicated by a particular $\chi^2_{lc}$ value is
dependent on the number of points in the light curve. We therefore converted the
$\chi^2_{lc}$ values into a probability that the given variation in the light
curve is statistically insignificant given the errors on each flux
measurement. This probability was calculated as the survival function for a
$\chi^2$ distribution with $n-1$ degrees of freedom. The probability of
variability was then converted to a significance level expressed in $\sigma$.
This approach breaks the degeneracy between degree and significance of
variability, which is not accounted for in early studies of variability, but
is recently becoming more common \citep[e.g,][]{bell_search_2015}. From the
analysis above we identified 86 sources as being variable at the $3\sigma$
level.

Due to the large number of false positive detections in the individual epochs,
each of the 86 candidate variable sources were manually inspected. The
following criteria were used to identify sources that are not considered to be
true variables:

\begin{enumerate}
\item the source is likely the side-lobe of a brighter source,
\item the source is a component of an extended or resolved source which is characterized by multiple components,
\item the source is only detected at the extreme edges of an image, or
\item the source is coincident with imaging artefacts that change between epochs.
\end{enumerate}

After manual inspection all but 9 sources were eliminated from the sample of
candidates. Of these 9 sources, 8 were detected in multiple epochs, whilst
one was detected in only one epoch, and is thus classified as a transient
source. These 9 sources are summarized in Table\,\ref{tab:sources} and
discussed in the following section.

\begin{table*}
\centering
\begin{tabular}{ccrrrcccc}
\hline
\hline
ID & RA/Dec & PDS Flux     & \multicolumn{2}{c}{Modulation Index}& Significance & Epochs & $\alpha_{0.843}^{1.4}$ & Notes\\
   & (J2000)  & (\mjy/beam)  &  m & m$_d$  & $\sigma$ \\
\hline
\hline
\multicolumn{6}{c}{Variable} \\
A & 01:08:21$-$45:28:32 & $84.7\pm  0.9$ & 0.18 & 14.4 &  6.9 & 4/4 & -1.3 & SUMSS J010821-452835\\
B & 01:09:53$-$46:31:29 & $19.8\pm  0.2$ & 0.16 & 10.2 &  3.3 & 3/3 & -1.2 & SUMSS J010952-463130\\
C & 01:10:14$-$46:15:07 & $13.5\pm  0.2$ & 0.07 &  2.8 &  3.0 & 5/5 & -2.2 & SUMSS J011015-461502\\
D & 01:11:14$-$45:36:01 & $ 6.7\pm  0.1$ & 0.13 &  6.9 &  4.6 & 6/6 & -1.0 & SUMSS J011114-453555\\
E & 01:12:17$-$46:29:32 & $ 6.8\pm  0.1$ & 0.19 & 11.8 &  5.0 & 5/5 & -    & \\
F & 01:13:40$-$46:03:47 & $15.7\pm  0.2$ & 0.09 &  5.4 &  7.4 & 5/5 & -0.8 & SUMSS J011341-460353\\
G & 01:14:10$-$46:35:48 & $48.7\pm  0.5$ & 0.08 &  3.9 &  4.9 & 3/3 & -0.4 & SUMSS J011410-463551\\
H & 01:15:44$-$45:55:50 & $39.8\pm  0.4$ & 0.11 &  6.5 & $>$8 & 3/3 & -0.8 & SUMSS J011544-455549; z=0.104 \\
\hline
\multicolumn{6}{c}{Transient} \\
T & 01:13:35$-$46:11:13 & $ 0.266\pm  0.044$ & 1.28 & 56.0 &  3.0 & 1/5 & - & \\
\hline
\hline
\end{tabular}
\caption{The 8 variable sources and one transient source with a significance greater than $3\sigma$. The reported flux is taken from \citet{hopkins_phoenix_2003}, and is equal to a weighted sum of the flux of a source across all epochs. The modulation indexes m and m$_d$ are calculated using equations \ref{eq:m} and \ref{eq:md}. The epochs column shows the number of epochs in which the source was detected and the total number of epochs in which the source could have been detected. The spectral index is taken between the PDS fluxes listed here and the corresponding measurement in the SUMSS catalog \citep[where it exists,][]{Mauch_sumss_2003}.}.
\label{tab:sources}
\end{table*}

\subsection{Variability}

The areal density of variable sources ($\rho$) is calculated as:
\begin{equation}
\rho = \frac{\Sigma N_V}{\Sigma A(N_E-1)}, \label{eq:eventrate}
\end{equation}

\noindent where $N_V$ is the number of variables, $N_E$ is the number of
epochs in which the source was observed, and $A$ is the area of sky covered by
$N_E$ epochs (see Figure~\ref{fig:coverage}). The summation is done over all
overlap regions to obtain a single source density for this work. We use the
above definition of variable source density throughout this paper. \citet{Mooley_caltech-nrao_2016}, and the associated online
resource\footnote{\href{http://www.tauceti.caltech.edu/kunal/radio-transient-surveys}{www.tauceti.caltech.edu/kunal/radio-transient-surveys}}, was used as a reference for calculating all variable source densities quoted in this paper.

The fraction of sources that is variable is also calculated over all overlap regions, using:
\begin{equation}
V(\%) = \frac{\Sigma N_V}{\Sigma N_S}, \label{eq:fractional}
\end{equation}
\noindent where $N_S$ is the number of sources detected within the region of sky after correction for reliability. The mean sensitivity of the observation is weighted by the area of each overlap region via:
\begin{equation}
\bar{\sigma} = \frac{\Sigma \sigma A}{\Sigma A}, \label{eq:sensitivity}
\end{equation}
\noindent where $\sigma$ is the mean sensitivity listed in Table\,\ref{tab:overlaps}.

Using the above equations, the data in Table~\,\ref{tab:sources} represent an
areal source density of variables of $\rho=0.86\,\mathrm{deg}^{-2}$, with a
sensitivity of $\bar{\sigma}=1.4$\mjy. The fractional of variable sources is
$\sim 0.7\%$ at 1.4\ghz{} on timescales of 6\,months to 8\,years. This fraction
is in agreement with that found by \citet{Mooley_sensitive_2013}, at the same
frequency but lower flux densities, in the extended Chandra deep field south.
We therefore support the suggestion of \citet{Mooley_sensitive_2013}, that the
1.4\ghz{} radio sky is relatively quiet.

\subsection{Transients}\label{subsec:transients}

One of the sources identified in Table\,\ref{tab:sources} was detected in only
a single epoch out of a possible five, and is thus a transient source. Using
the same method as the previous section, we find a density of transients $\rho
= 0.1\,\mathrm{deg}^{-2}$ with a sensitivity of $\bar{\sigma}=1.4$\mjy. Given
a single transient source across all epochs, we estimate that in any image
$0.1\%$ of all point sources above $1.4$\mjy/beam will be a transient source.
This density is consistent with an  extrapolation from other studies at
$1.4$\,GHz \citep{Mooley_sensitive_2013,croft_allen_2010,croft_allen_2011}.

\begin{figure}
\centering
\includegraphics[width=\linewidth]{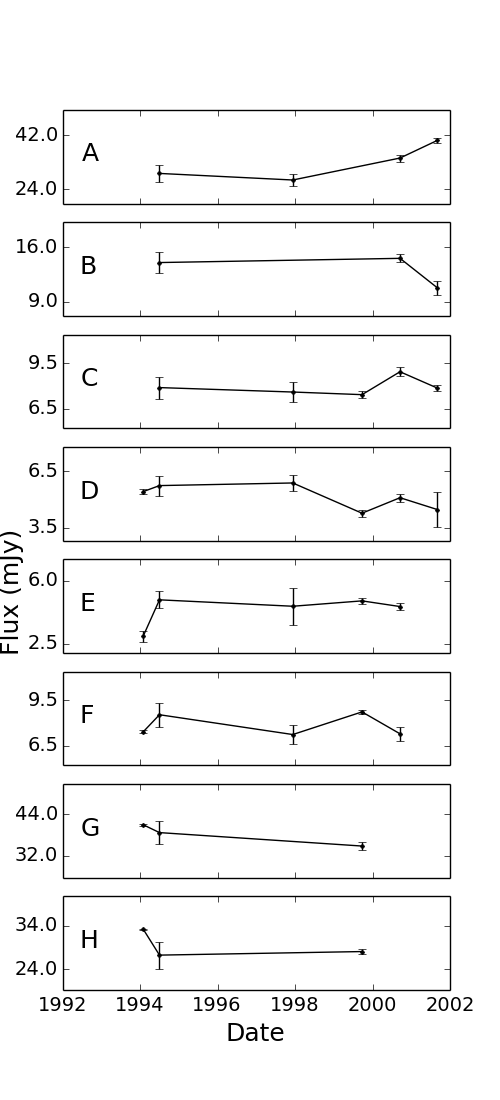}
\caption{The light curves of variable sources A-H, scaled to emphasize the significance of variability. Error bars show $1\sigma$ uncertainties.}
\label{fig:stackedlc}
\end{figure}

\section{Variable Source Analysis}\label{sec:variable_source_analysis}

The light curves of all the variable sources are shown in
Figure\,\ref{fig:stackedlc}. The sparse sampling of these light curves
precludes any indication of the cause of variability so we turn instead to
multi-wavelength data to understand the nature of the sources and possibly the
cause of variability.

Each of the sources were cross-matched with the SUMSS catalog
\citep{Mauch_sumss_2003} and for all but sources E and T, a counterpart was found.
The counterparts are all point sources and we calculate the average spectral
index for each using the flux from the PDS and SUMSS catalogs. The spectral
index of each source is shown in column 8 of Table~\ref{tab:sources}. We note
that the spectral indexes are all negative, which is in contrast to the
positive spectral indexes that \citet{bell_search_2015} measured for variable
sources found at 5\ghz{}. Unlike the recent observations by
\citet{bell_search_2015} the PDS observations were made prior to the ATCA
broad-band upgrade, and so we are not able to extract an intra-band spectral
index for any of the sources in our sample. A steep (or at least negative)
spectral index is consistent with the optically thin spectrum of an AGN.

\begin{figure}
\centering
\includegraphics[width=0.95\linewidth]{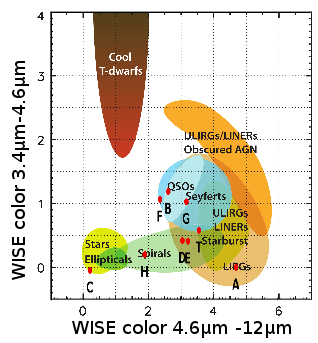}
\caption{Each of the variable sources are identified according to their WISE colors. The red points are labeled according to the source ID, whilst the colored regions indicate the source classification. This image modified from \citet{wright_wide-field_2010}.}
\label{fig:wise}
\end{figure}

Comparing the variable sources to the Wide-field Infrared Survey Explorer
\citep[WISE,][]{wright_wide-field_2010} catalog we are able to obtain a cross
identification for all sources. We place the sources on a color-color plot as
shown in Figure~\ref{fig:wise}. All the sources but two are consistent with an
AGN of some description. Source C has colors that are consistent with either a
star or elliptical galaxy, and source H is consistent with a spiral galaxy.

Sources A, C, and H were found to have matches in additional catalogs
hosted by VizieR\footnote{\href{http://vizier.u-strasbg.fr/viz-bin/VizieR}{vizier.u-strasbg.fr/viz-bin/VizieR}}
and are discussed in detail below, along with the transient source T.

\subsection{Source A}

This is a previously known radio source that has been detected in a number of
other radio surveys, as summarized in Table\,\ref{tab:srcA}. The data are well
described by a single power law with a spectral index of $\alpha=-0.77$.
Source A is identified by \citet{Flesh_all-sky_2010} as having an X-ray
counterpart, and they assign a 93\% probability that this source is a quasar. The
spectral index of this source is consistent with an optically thin synchrotron
emission. The variability that is observed could be due to intrinsic
variability in the fueling of an AGN. With a de-biased modulation index of
just 14.4 over a timescale of less than 8 years, intrinsic variability is
certainly possible.

\begin{table}
\centering
\begin{tabular}{cll}
\hline
Frequency & Flux & Reference \\
GHz & mJy & \\
0.180 & $588\pm 73$   & MWACS; \citet{Hurley-walker_murchison_2014}\\
0.843 & $160\pm 0.49$ & SUMSS; \citet{Mauch_sumss_2003}\\
1.4   & $84.7\pm 0.9$ & PDS;   \citet{Hopkins1998}\\
4.8   & $49\pm 10$    & PMN;   \citet{Gregory_parkes_1994}\\
\hline
\end{tabular}
\caption{The flux of source A as measured in multiple radio surveys. The data are well described by a single power law with spectral index of $\alpha=-0.77$.}
\label{tab:srcA}
\end{table}

\subsection{Source C}

\citet{afonso_phoenix_2005} identify this source as a star with an r-magnitude
of 14.89. This designation is consistent with the WISE colors seen in
Figure~\ref{fig:wise}. Source C appears in the XMM serendipitous source
catalog \citep[3XMM-DR4,][]{Rosen_3XMM-DR4_2015} with a soft spectrum. The
photon count in the XMM catalog is too low to obtain a secure classification
using just the X-ray data. However when taken in combination the radio, infra-red and X-ray data for source C are consistent with a massive hot star with an
unstable wind \citep{Kudritski_winds_2000}.

\subsection{Source H} 

The infrared colors of this source indicate that the host galaxy is a spiral
(Figure~\ref{fig:wise}). \citet{afonso_phoenix_2005} observed this source as
part of the PDS follow up and reported a redshift of $z=0.104$ with a spectrum
that indicates star-formation and narrow emission-line system.
\citet{afonso_phoenix_2005} also report a $H_\alpha$ luminosity of
$10^{34.83}W$ and 1.4\ghz{} luminosity of $10^{24.3}W\,Hz^{-1}$, making it the
most radio luminous of all the star-forming galaxies identified in the follow
up observations. Variability in a galaxy with star formation indicates that
there are multiple sources of emission, with only the compact component
being variable. Narrow emission lines are common but not exclusive to in AGN,
however the presence of radio variability is evidence for an AGN core. We
conclude that this source is an AGN with active star-formation.

\subsection{Source T}

\begin{figure}
\centering
\includegraphics[width=0.95\linewidth]{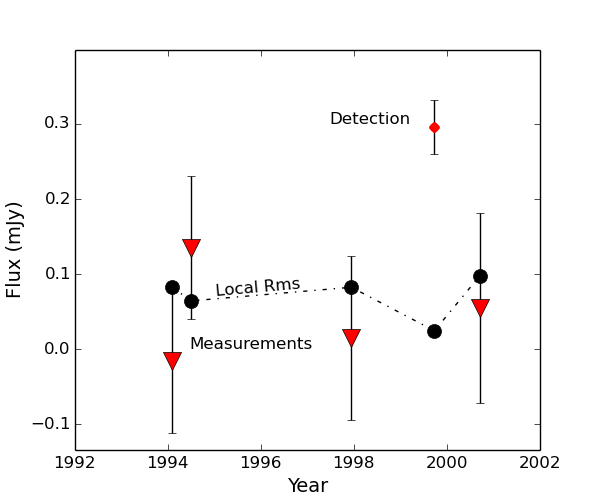}
\includegraphics[width=0.95\linewidth]{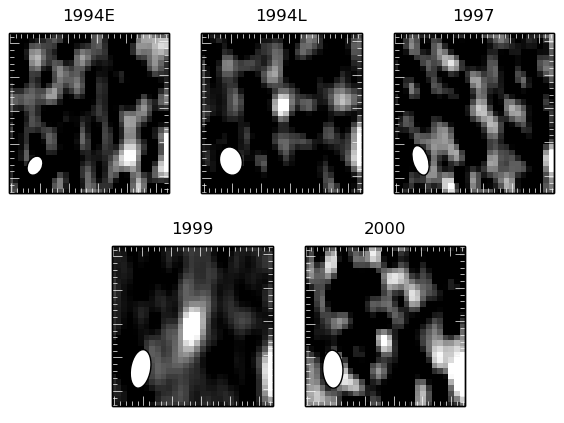}
\caption{Upper: The light curve for the transient source T. The single detection is the highest labeled point, and the triangular points show the measurements in the remaining epochs. The dot-dashed line and points indicated the $1\sigma$ local noise at each epoch. Lower: Images centered on the location of the transient (01:13:35$-$46:11:13) for each of the five epochs that covered this region of sky. The white ellipse is the size of the synthesized beam, and the images are 0\farcm 48 on a side.}
\label{fig:transient}
\end{figure}

As mentioned in the previous section, source T was identified in the
variability search and was detected in only a single epoch out of a total of
five observations. The flux at the location of the transient was measured in
the remaining four observations. In three epochs (1994E, 1997, 2000) the
measurements is consistent with zero flux, however in epoch 1994L the
measurement indicates a source with a non-zero flux but at the $2\sigma$
level. Figure\,\ref{fig:transient} shows the light curve for the transient
source, along with a sequence of images from each of the five epochs of
observations.

The 1999 epoch detection is $12.8$ times the local rms, making this single
detection highly significant, even though the significance of the variability
is only $3\sigma$. The PDS catalog lists the flux of this source as being
$0.266\pm 0.044$\mjy, whilst the 1999 epoch detection is at a flux of
$0.314\pm 0.034$\mjy. This agreement in flux is due to the fact that at the
location of source T the 1999 image has a very small local noise
($0.025$\mjy/beam), where as the other epochs have a local rms that is 3-5 times
greater. The linear mosaicking that was used by \citet{Hopkins1998} used a
weighting scheme that was proportional to the inverse square of the local rms
and thus the signal from the 1999 epoch dominates the flux measurement.

The fact that source T has a low significance detection in the 1994L epoch
suggests that the source may have some amount of quiescent emission. It is
possible that we are seeing a faint AGN that is undergoing interstellar
scintillation and that in the 1999 epoch scintillation has boosted the flux of
the source to a detectable level. If this is the case then source T would have
a light curve more similar to the variable sources that have been previously
discussed. An infrared counterpart was detected for this source, and the
counterpart has colors consistent with an AGN (see Figure~\ref{fig:wise}). The
light-curve and the designation as an AGN means that although this sources was
detected as a transient we consider this to be a variable source at the edge
of our detection limit. If we consider this source to be a variable rather
than a transient, then we arrive at a revised source density of
$\rho=0.98\,\mathrm{deg}^{-2}$ for variables, and an upper limit of
$<0.1\,\mathrm{deg}^{-1}$ for transients.

\section{Discussion}\label{sec:discussion}

\begin{figure*}
\includegraphics[width=\linewidth]{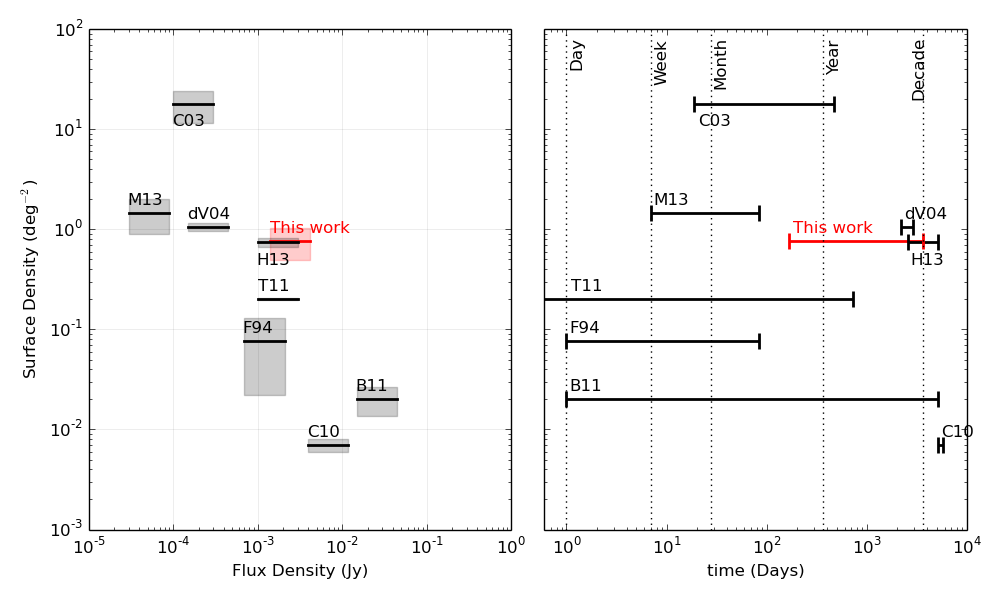}
\caption{Two epoch equivalent areal density of variable sources detected at 1.4GHz. Left: Source density as a function of flux density. Shaded boxes represent Poisson errors on the number of detected variable sources. Right: The timescales probed by the surveys indicated in the left panel. The variables found within the phoenix field have an areal density that agrees with that measured by \citetalias{Hodge_millijansky_2013}. The range of timescales probed by this work bridges a gap between the timescales probed by previous surveys. References are: C03 - \citep{carilli_variability_2003}, M13 - \citep{Mooley_sensitive_2013}, dV04 - \citep{deVries_optical_2004}, H13 - \citep{Hodge_millijansky_2013}, F94 - \citep{Frail_search_1994}, C10 - \citep{croft_allen_2010}, B11 - \citep{bannister_22-yr_2011}, T11 - \citep{thyagarajan_variable_2011}, and this work.}
\label{fig:lognlogs}
\end{figure*}

Figure\,\ref{fig:lognlogs} compares the areal source density found in this
work to that of other radio variability studies at $\sim 1$\,GHz. In compiling
data for Figure\,\ref{fig:lognlogs}, we consider only blind radio surveys for
variability at frequencies between 0.5 and 2\,GHz. The timescales of
variability that are probed by the various studies are also shown in
Figure\,\ref{fig:lognlogs}, for comparison. If we consider all the 9 sources
identified in the previous section as variables, then we measure a variable
areal source density of $\rho=0.98\mathrm{deg}^{-2}$. If we look at surveys
with a sensitivity of $\sim 1$\mjy/beam we have: \citet{Frail_search_1994}
($\rho=0.076\,\mathrm{deg}^{-2}$) on timescales of days to months,
\citetalias{thyagarajan_variable_2011} ($\rho=0.2\,\mathrm{deg}^{-2}$) on
timescales of minutes to years, this work on timescales of 6 months to 8
years, and finally \citetalias{Hodge_millijansky_2013}
($\rho=0.74\mathrm{deg}^{-2}$) on timescales of $7-14$ years. There is a
common story unfolding here: surveys with a cadence of years-decades detect
more variable sources than those with a cadence of days-months. It should be
noted that the work of \citet{croft_allen_2010} probed the longest time scales
of all the surveys in Figure\,\ref{fig:lognlogs}, but found a source density
that is the lowest of all. This would suggest that the relationship between
observing cadence and variable source density is not monotonic. Both the
increase of variability on longer timescales, and the decrease at the longest
timescales is consistent with the suggestion of \citet{bannister_22-yr_2011}
that variability is greatest on timescales of $\sim 2-5$\,yr.

A peak in the variability of radio sources as a function of observing cadence
is in disagreement with the work of \citet{ofek_structure_2011} who measure a
structure function of variability is flat. However \citet{ofek_structure_2011}
rely on data that spanned timescales of 1 month to 5 years which is both
longer than the shortest timescales accessible to
\citetalias{thyagarajan_variable_2011}, and shorter than the shortest
timescales probed by \citetalias{Hodge_millijansky_2013}. Despite this
disagreement on the incidence of variability, we agree with
\citet{ofek_structure_2011} on the primary cause of variability - interstellar
scintillation of a compact component making up at least a fraction of the flux
observed in each of the sources.

Of the 9 identified sources, all but one have a counterpart in the SUMSS
survey and a negative spectral index. These spectral indexes are consistent
with classical AGN in the optically thin regime. If we consider intrinsic
variability with an AGN then we would expect the light curve to be a
superposition of self absorbed SEDs each cooling and passing through the
observing band. Thus the spectral index would be positive and negative for
approximately equal amounts of time. In our snapshot of 9 sources we find only
negative spectral indexes which argues against such a model of intrinsic
variability. Alternatively, we suggest that the variability that we are
primarily observing is interstellar scintillation of AGN
\citep{ofek_structure_2011}.

Source C is a notable exception to the above arguments in that it is not an
AGN. In this case the emission is coming from shocks in the wind of a hot
star \citep{Kudritski_winds_2000}, on scales much less than a light year.
Intrinsic variability on timescales of years has been seen before in such
objects \citep[eg,][]{Loo_non-thermal_2008} and is thus a possible
explanation.

For extragalactic radio sources, we find that radio variability is dominated by interstellar scintillation.
The scintillating medium is necessarily of Galactic origin.
Investigations into long-term radio variability at $\sim 1\ghz$ frequencies tell us less about the observed sources, and more about the nature of gas within our own Galaxy.
This focus is contrary to the stated goals of many radio variability surveys.
Future radio surveys which aim to explore a new parameter space of transient and variable objects, with a focus on intrinsic variability,  should thus focus attention on short timescales (and thus explosive events).
However, long-time scale variability studies can offer a new insight into the distribution and behavior of gas within the Milky Way.

\section{Summary and Conclusions}\label{sec:summary}  

We used six epochs of data from the Phoenix Deep Survey to search for
variables and resolve the conflict between the results of
\citet{thyagarajan_variable_2011} and \citet{Hodge_millijansky_2013}. We
measure the density of variable radio sources to be
$\rho=0.98\mathrm{deg}^{-2}$ which is consistent with that of
\citet{Hodge_millijansky_2013}. Given the overlap in timescales probed by this
work that and that of \citet{Hodge_millijansky_2013}, and the shorter timescales
probed by \citet{thyagarajan_variable_2011}, we suggest that the discrepancy in
variable source density could be due to the different time scales that were
probed. 

We have made use of infrared colors from the WISE survey to provide a fast identification of source types, and fond that all variable sources are consistent with an AGN.
This approach is likely to find continued use in large area transient and variable surveys, where automated host typing will allow for more appropriate follow up observations.
This will be particularly important for the search for hosts of Fast Radio Bursts and Gravitational Wave events.

Eight of the nine variable sources detected in this work show behavior
consistent with interstellar scintillation of AGN, whilst the remaining source
shows variability that can be attributed to intrinsic causes. Our results
support the claim of \citet{bannister_22-yr_2011}, that variability at $\sim
1\ghz$ frequencies is greatest on timescales of $2-5$ years.

\section*{Acknowledgments}   

We thank Andrew Hopkins and Jos\'e Afonso for provision of the calibrated uv data from the Phoenix survey.
We also thank the anonymous referee for their suggestions which led to the improvement of this work.

B.~M.~G. acknowledges the support of the Australian Research Council through
grant FL100100114. The Dunlap Institute is funded through an endowment
established by the David Dunlap family and the University of Toronto.

Parts of this research were conducted by the
Australian Research Council Centre of Excellence for All-sky Astrophysics
(CAASTRO), through project number CE110001020.

This publication makes use of data products from the Wide-field Infrared
Survey Explorer, which is a joint project of the University of California, Los
Angeles, and the Jet Propulsion Laboratory/California Institute of Technology,
funded by the National Aeronautics and Space Administration.

This research has made use of the VizieR catalogue access tool, CDS,
Strasbourg, France. The original description of the VizieR service was
published in A\&AS 143, 23.

This research has made use of data obtained from the 3XMM XMM-Newton
serendipitous source catalogue compiled by the 10 institutes of the XMM-Newton
Survey Science Centre selected by ESA.

\bibliographystyle{mn2e}
\bibliography{mybib}
\label{lastpage}
\end{document}